\documentclass[preprint,12pt]{elsarticle}

\usepackage{amssymb}
\usepackage{color,graphicx}
\usepackage{amsmath}
\usepackage{amsbsy}
\usepackage{amsthm}
\usepackage{bbm}
\usepackage{bm}
\usepackage{epsfig}
\usepackage{lscape}
\usepackage{float}
\usepackage{graphicx}
\usepackage{dcolumn}
\usepackage{bbm}
\usepackage{color,epstopdf}
\usepackage{amscd}
\usepackage{amsfonts}
\usepackage{amsmath}
\usepackage{amssymb}
\usepackage{verbatim}

\journal{Physics Letters A}

\begin{document}

\begin{frontmatter}

\title{Gravitational Decoherence for Mesoscopic Systems}

\author{Stephen L. Adler}
\address{Institute for Advanced Study, Einstein Drive, Princeton, NJ 08540, USA}

\author{Angelo Bassi}
\address{Department of Physics, University of Trieste, Strada Costiera 11,
34151 Trieste, Italy}
\address{Istituto Nazionale di Fisica Nucleare, Trieste Section, Via Valerio
2, 34127 Trieste, Italy}

\begin{abstract}
We extend the recent gravitational decoherence analysis of Pikovski et al. to an individual mesoscopic system with internal state characterized
by a coherent superposition of energy eigenstates.  We express the Pikovski et al. effect directly in terms of
the energy variance, and show that the interferometric visibility is bounded from below.  Hence unlike collisional decoherence,
the visibility does not approach zero at large times, although for a large system it can become very small.

\end{abstract}
\end{frontmatter}


In a recent  interesting article, Pikovski et al.~\cite{pik}  show that there is a universal decoherence-like effect for systems in a varying gravitational potential.
They focus on a system consisting of a large number $N$ of thermally excited harmonic oscillators, and calculate a formula, with which we agree, for
the time dependence of the interferometric visibility.  Their formula depends on the subsystem number and temperature $T$ as $N^{1/2}k_BT$ (with $k_B$ the Boltzmann
constant), and as they note this indicates
that the internal energy variance is the relevant system attribute for their effect, which is explicit in their earlier paper \cite{pik2} discussing gravitational decoherence in a two state system. Two recent papers \cite{bonder}, \cite{diosi} have noted that the Pikovski et al. effect vanishes in a freely falling frame, and so it furnishes yet another interesting example of the subtle interplay of quantum theory and gravitational physics.
In this paper, however, we confine ourselves to the frame of an asymptotically (with respect to the Earth) inertial observer; here the nonrelativistic Newtonian limit suffices
to give correct answers.

In this note we extend the  derivation of \cite{pik} to a single mesoscopic
system characterized by a general coherent superposition of energy eigenstates, and show directly that the gravitational reduction in interferometric visibility depends,
for small times $t$,  on the energy variance $\Delta E$.  We also show that the visibility is strictly bounded from below, and does not approach
zero for large times.  When extended to a large collection of $N$
independent mesoscopic subsystems, with thermal energy variance $\Delta E \sim N^{1/2}k_BT$, our result maps to that of Pikovski et al., and the lower bound vanishes exponentially for large $N$.  However, this exponential vanishing is critically dependent on the independence assumption,
and may not apply to strongly coupled subsystems.

To keep this article self-contained, we present a complete derivation, starting from the underlying gravitational physics.
According to general relativity, gravity couples to matter via the stress-energy tensor $T_{\mu\nu}({\bf x})$; in the non-relativistic limit and to first order in weak gravitational fields, the interaction Hamiltonian is
\begin{equation}
H_{\text{\tiny G}} = \frac{1}{c^2} \int d^3 x \phi({\bf x}) T_{00} ({\bf x})~~~,
\end{equation}
where $\phi({\bf x})$ is the Newtonian potential. If the energy density is well localized in space with respect to the distances over which $\phi({\bf x})$ varies appreciably, we can bring $\phi({\bf x})$ outside the integral to obtain $H_{\text{\tiny G}} = \phi(\bar{\bf x}) H / c^2$, where $\bar{\bf x}$ is the mean position of the matter content and $H$ is the {\it total} Hamiltonian of the matter distribution. In the case of the Earth's gravitational field, where $\phi(x)=gx$, with $g$ the Earth's gravitational acceleration and
$x$ the vertical distance above Earth's surface, we therefore have
\begin{equation}
H_{\text{\tiny G}} = g \bar x H / c^2~~~.
\end{equation}
The total $H$ is given by
\begin{equation} \label{eq:ham}
H = mc^2 +H_{\text{\tiny center-of-mass~kinetic}}+ H_{\text{\tiny int}}~~~,
\end{equation}
with $H_{\text{\tiny int}}$ the system internal energy.
The dominant term is the rest mass $mc^2$, which produces the standard Newtonian term $V = m g \bar x$ in the Schr\"odinger equation, the quantum effects of which have been measured in neutron interferometry \cite{cow}.
The  internal energy contribution in Eq. \eqref{eq:ham} introduces new effects, specifically a phase shift in the time evolution of the internal energy states, which is the origin for the time dilation decoherence discussed by Pikovski et al.  In the calculation that follows, we neglect the $mc^2$ term in the energy, as well as the  kinetic energy associated with center of mass motion, since these do not contribute to the internal energy variance and so drop out of the final formula.

Let us now consider a single system and calculate for this the analog of the Pikovski et al. effect.
We assume that the total initial state of the system is $| \psi(x, 0) \rangle = | \psi_{\text{\tiny CM}}(x) \rangle \otimes | \psi_{\text{\tiny int}}\rangle$, where $| \psi_{\text{\tiny CM}}(x) \rangle$ is the center-of-mass wave function, which is assumed to be well-localized in space around the position $x$, while $| \psi_{\text{\tiny int}}\rangle$ is the internal dynamics, which we decompose as a superposition of eigenstates $|n\rangle$ of the internal Hamiltonian: $| \psi_{\text{\tiny int}}\rangle = \sum_n c_n |n\rangle$. The state at time $t$ is thus
\begin{eqnarray}\label{eq:state}
| \psi(x, t) \rangle = &| \psi_{\text{\tiny CM}}(x) \rangle \otimes | \psi_{\text{\tiny int}}(x,t)\rangle~~~,\nonumber \\
| \psi_{\text{\tiny int}}(x,t)\rangle =&  \sum_n c_n e^{-i E_n t (1 + gx/c^2 )/\hbar} |n\rangle~~~.
\end{eqnarray}
We see that due to the presence of the Earth's gravitational field, the internal states acquire a phase which depends on the position of the center of mass with respect to the Earth.   For a single system  in an energy eigenstate, the gravitational effect appears only as an overall phase of the internal wave function and does not change its magnitude; the more interesting case is when two or more energy eigenstates appear
in the sum over $n$ in Eq. \eqref{eq:state}, so that the gravitational effect cannot be factored out as an overall phase.  Note that even
in this more complex case, the gravitational effect is reversible, in that if the system is sent through a gravitational field, and then
through an equal but reversed gravitational field, the effect is eliminated.  This already indicates a contrast with standard collisional  decoherence induced by interaction with a chaotic environment, which cannot for all practical purposes be reversed.

  We now consider a system that is in a superposition of two different positions in space, $| \psi_{\text{\tiny CM}}(x) \rangle \rightarrow [| \psi_{\text{\tiny CM}}(x_1) \rangle + | \psi_{\text{\tiny CM}}(x_2) \rangle] /\sqrt{2}$, with the two center-of-mass states  well localized with respect to $\Delta x = x_1 - x_2$, and practically orthogonal.  Then the internal states will entangle with the center of mass and the reduced density matrix $\hat{\rho}^{\text{\tiny CM}}$ of the center of mass, obtained by tracing the density matrix over the internal degrees of freedom, will dephase. More precisely, the off-diagonal element $\rho^{\text{\tiny CM}}_{12} = \langle \psi_{\text{\tiny CM}}(x_1) | \hat{\rho}^{\text{\tiny CM}} | \psi_{\text{\tiny CM}}(x_2) \rangle$ evolves in time as follows,
\begin{equation} \label{eq:cm}
\rho^{\text{\tiny CM}}_{12} (t) =  \frac{1}{2}\sum_n |c_n|^2 e^{-i E_n t g \Delta x  / \hbar c^2}.
\end{equation}
We will be interested in calculating the interferometric visibility, which here is twice the absolute value $|\rho^{\text{\tiny CM}}_{12} (t)|$.  This is unchanged when we multiply
$\rho^{\text{\tiny CM}}_{12} (t)$ by any phase factor, and it is convenient to choose this phase factor as follows.
Let us define $\bar{E} =  \sum_n |c_n|^2 E_n$ as the mean  internal energy, and define  $\Delta E_n = E_n - \bar{E}$, so that   $\sum_n |c_n|^2 \Delta E_n = 0$ and $(\Delta E)^2 = \sum_n |c_n|^2 (\Delta E_n)^2$ gives the square of the internal energy variance $\Delta E$.  Then we get the same absolute value $|\rho^{\text{\tiny CM}}_{12} (t)|$ if we
 replace $\rho^{\text{\tiny CM}}_{12} (t)$  by $\tilde{\rho}^{\text{\tiny CM}}_{12} (t) = e^{i \bar{E} t g \Delta x/\hbar c^2} \rho^{\text{\tiny CM}}_{12} (t) $.

Expanding $ 2\tilde{\rho}^{\text{\tiny CM}}_{12} (t)$   in powers of $t$, we get
\begin{eqnarray}\label{eq:expansion}
2\tilde{\rho}^{\text{\tiny CM}}_{12} (t) & = &  \sum_n |c_n|^2 \left[ 1 - i g \frac{\Delta E_n  \Delta x }{\hbar c^2} t - \frac{1}{2} g^2  \frac{(\Delta E_n)^2 { \Delta x}^2 }{\hbar^2 c^4} t^2 +  O(t^3) \right] \nonumber \\
& = & 1 - t^2/t_{\text{\tiny D}}^2 + O(t^3),
\end{eqnarray}
where we have introduced the phase evolution time scale (which governs the Pikovski et al. effect),
\begin{equation} \label{eq:td}
t_{\text{\tiny D}} = \frac{\sqrt{2} \hbar c^2}{g \Delta E |\Delta x|}.
\end{equation}
The phase evolution time depends on the spatial separation of the two center of mass components as well as on the internal energy variance, and the behavior of Eq. \eqref{eq:cm} depends strongly on the system complexity.  For a single system consisting of a superposition of only two energy
eigenstates, $2|\tilde{\rho}^{\text{\tiny CM}}_{12} (t)|$ oscillates in time, on a time scale of order of magnitude $t_D$.  For a
single system consisting of a superposition of many eigenstates, over large time scales the behavior of the visibility is still oscillatory, and is governed by the lower bound derived below.  But for  a complex system the behavior of the interferometric visibility for small times $t \ll t_D$ is of interest in understanding the Pikovski et al. effect, and is given by
\begin{equation}
2|\rho^{\text{\tiny CM}}_{12} (t)|\simeq  1 - t^2/t_{\text{\tiny D}}^2~~~,
\end{equation}
and for a system composed of $N$ identical subsystems, the visibility for small times is given by
\begin{equation}\label{eq:exp1}
2|\rho^{\text{\tiny CM}}_{12} (t)|\simeq  (1 - t^2/t_{\text{\tiny D}}^2)^N \simeq \exp(-N t^2/t_{\text{\tiny D}}^2) ~~~.
\end{equation}
This can be rewritten as
\begin{equation}\label{eq:tgrav}
2|\rho^{\text{\tiny CM}}_{12} (t)|\simeq \exp(- t^2/t_{N\text{\tiny D}}^2) ~~~,
\end{equation}
with
\begin{equation} \label{eq:td1}
t_{N\text{\tiny D}} = \frac{\sqrt{2} \hbar c^2}{\sqrt{N}g \Delta E |\Delta x|},
\end{equation}
in which the effective energy variance $\Delta E$ is multiplied by $\sqrt{N}$.  Note that the exponentially vanishing behavior of Eq. \eqref{eq:exp1} for large $N$ is critically dependent on the assumption of $N$ {\it independent} systems.  For a strongly coupled $N$ particle  system with $\Delta E_N \sim N^{1/2} \Delta E_1$,  without further
input we could only conclude that the right hand side of Eq. \eqref{eq:exp1} is replaced by $1-N t^2/ t_{1D}^2$ for small $t$ \big(i.e. Eq. \eqref{eq:expansion} with $\Delta E$ replaced by $\Delta E_N$, the total internal energy variance\big), which does not imply
exponentially vanishing behavior of the visibility for large $N$.  A similar cautionary remark applies to the lower bound on the visibility of a system composed of $N$ independent
subsystems derived in Eq. \eqref{eq:lower} below.

If the initial internal state is not a pure state, but a statistical mixture of states $| \psi_{\text{\tiny int}}^{\alpha}\rangle = \sum_n c_n^{\alpha} |n\rangle$ with probabilities $p_{\alpha}$, then Eq.~\eqref{eq:td} still holds, with:
\begin{equation}
(\Delta E )^2 = \sum_{\alpha, n} p_{\alpha} |c_n^{\alpha}|^2 (\Delta E_n)^2,
\end{equation}
which measures internal energy fluctuations both of `quantum' origin due to the eigenstate superposition  with amplitudes $c_n$, as well as  of `classical' origin due to the statistical  probabilities  $p_\alpha$.  Pikovski {\it et al.} consider the case of only thermal fluctuations, in which case $\Delta E = \sqrt{N} k_B T$, where $N$ is the number of degrees of freedom, $k_B$ Boltzmann's constant and $T$ the temperature, and one recovers Eq. (3) of their paper.  This also agrees with Eq. \eqref{eq:td1} above for the case of $N$ independent subsystems with
individual energy variance $\Delta E =k_B T$.

We now show that the effect produced by the coupling of the internal degrees of freedom to the gravitational field, being a sum of phase shifts, does not correspond to decoherence in the usual sense. Forming the absolute value squared of the visibility, we have
\begin{equation} \label{eq:cm2}
(2|\rho^{\text{\tiny CM}}_{12} (t)|)^2 =  \sum_{n,m} |c_n|^2 |c_m|^2 e^{-i (\Delta E_n - \Delta E_m) t g \Delta x  / \hbar c^2},
\end{equation}
which implies that the average of the visibility over a large time interval is
\begin{equation}\label{eq:bound}
\lim_{T \rightarrow +\infty} \frac{1}{T} \int_0^T (2|\rho^{\text{\tiny CM}}_{12} (t)|)^2 dt = \sum_n |c_n|^4 > 0,
\end{equation}
where we have assumed $\Delta E_n \neq \Delta E_m$ for $n \neq m$.    When there are energy degeneracies, this formula still holds, with $|c_n|^2$ the sum of the absolute value squared coefficients of all
states with the same energy.

 Since $ \sum_n |c_n|^2=1$, the lower bound of Eq. \eqref{eq:bound} is always a number smaller than 1, but greater than 0.  A stronger statement
 can be made when the number of states in the superposition is a finite number $L$.  Expanding out the inequality
 \begin{equation}\label{eq:llevel}
 \sum_{m=1}^L\sum_{n=1}^L (|c_n|^2-|c_m|^2)^2 \geq 0~~~,
 \end{equation}
 and using state vector normalization  $\sum_{n=1}^L |c_n|^2=1$ and  state counting $\sum_{n=1}^L 1=L$ gives
 \begin{equation}\label{eq:llevel1}
 \sum_{n=1}^L  |c_n|^4 \geq L^{-1}~~~.
 \end{equation}
 Equations \eqref{eq:bound} and \eqref{eq:llevel1} mean that $| \rho^{\text{\tiny CM}}_{12} (t)|$ cannot vanish for large times. This is in contrast to usual collisional decoherence, where the effect of a single interaction is of the form $\rho^{\text{\tiny CM}}_{12} \rightarrow c_{12} \rho^{\text{\tiny CM}}_{12}$ with $|c_{12}| < 1$, and so after $N_i$  interactions, the off-diagonal element approaches zero as
$|c_{12}|^{N_i}$. When $N_i$ is linear in time this gives an exponential vanishing of the visibility as a function of time.  By way of contrast, the gravitational decoherence calculated
above gives a visibility that vanishes exponentially as the number of independent subsystems $N$ approaches infinity, but which has a Gaussian decrease in time only for small times.
For $N$ independent subsystems, the lower bound of Eq. \eqref{eq:bound} is replaced by
\begin{equation}\label{eq:lower}
(\sum_n |c_n|^4)^N~~~,
\end{equation}
which since $ \sum_n |c_n|^4<1$ approaches zero exponentially as the subsystem number $N$ approaches infinity.

As a concrete illustration of the smallness of the lower bound for large systems, consider a cube $10^{-7}$ cm on a side, containing roughly
1000 atoms, with of order 1000 acoustical vibration modes.  Since these modes are approximately independent, we can apply Eq. \eqref{eq:lower};
assuming  that the cube state is prepared so that the average mode is in a superposition of 3 states, we can also use Eq. \eqref{eq:llevel1} with $L=3$ as an estimate for $\sum_n|c_n|^4 $, giving for the lower bound on the long time average of the visibility
\begin{equation}\label{eq:estimate}
\lim_{T \rightarrow +\infty} \frac{1}{T} \int_0^T (2|\rho^{\text{\tiny CM}}_{12} (t)|)^2 dt \geq 3^{-1000},
\end{equation}
which is zero for all practical purposes.  With this same model in mind, it is also instructive to consider the competition between
gravitational decoherence and standard collisional decoherence.  For collisional decoherence, in the limit of small spatial superpositions, the analog of Eq. \eqref{eq:tgrav} is \cite{coll}
\begin{equation}\label{eq:tcoll}
|\rho_{12}^{\rm CM}(t)| = |\rho_{12}^{\rm CM}(0)|\exp(-t/t_{\rm Coll})~~~,
\end{equation}
with
\begin{align}\label{eq:tcolldef}
t_{\rm Coll}= &\frac{1}{\Lambda |\Delta x|^2}~~~,\cr
\Lambda =&\frac{n \sigma \langle q^2v\rangle_{\rm AV}}{3\hbar^2}~~~.
\end{align}
In this formula, $n$ is the density of scatterers that scatter from the decohering object with cross section $\sigma$, and $q$ and
$v$ are respectively the scatterer momentum and velocity.  If the scattering particles have mass $m$ and form a thermal bath at
temperature $T$, then
\begin{equation}\label{eq:q2vav}
\langle q^2v\rangle_{\rm AV}=4 (m/\pi)^{\frac{1}{2}} (2 k_{\rm B}T)^{\frac{3}{2}}~~~.
\end{equation}
Assuming that the decohering object is in thermal equilibrium with the bath, we can take its
temperature also as $T$, and so in the formula for $t_{N_D}$ we can take $\Delta E = \sqrt{N} k_B T$.  Combining the various formulas,
the condition for $t_{N_D} < t_{\rm Coll}$ can be written as
\begin{equation}\label{eq:condition}
n\leq \frac{3(N\pi)^{\frac{1}{2}}}{16}\frac{\hbar g}{c^2 |\Delta x| \sigma (m k_B T)^{\frac{1}{2}} }~~~.
\end{equation}
Taking as an example the 10 atom cube and assuming the scatterers are nitrogen molcules at room temperature, we have as inputs for
a numerical estimate of decoherence of a superposition in which the cube center is displaced by the cube diameter,
\begin{align}\label{eq:inputnumbers}
\sigma=&10^{-14}{\rm cm}^2  ~~~,\cr
N=&1000~~~,\cr
k_B T = & \frac{1}{39} \rm{eV}~~~,\cr
|\Delta x| =& 10^{-7} {\rm cm}~~~,\cr
m = & 14 \times 10^9 {\rm eV}/c^2 ~~~,\cr
g = &981 {\rm cm}/{\rm s}^2 ~~~,\cr
\hbar= &6.6 \times 10^{-16} {\rm eV}{\rm s}~~~,\cr
c=&3 \times 10^{10} {\rm cm}/{\rm s}~~~,\cr
\end{align}
the inequality of Eq. \eqref{eq:condition} becomes
\begin{equation}\label{eq:final}
n \leq 1.2 \times 10^{-5} {\rm cm}^{-3}~~~.
\end{equation}
This is a density of $10^{-24}$ of atmospheric density at standard temperature and pressure, and corresponds to a vacuum
presently unattainable in the laboratory.   So under normal laboratory conditions, collisional decoherence occurs on a more
rapid time scale than gravitational decoherence.

To summarize, we have given a generalized and simplified derivation of the gravitational decoherence effect of \cite{pik} that applies to a single mesoscopic system containing an arbitrary superposition of energy states, and have shown that there are important limitations on the size and applicability of the effect.

The authors wish to thank Edward Witten and Hendrik Ulbricht for helpful discussions.  A.B. wishes the acknowledge the hospitality of the Institute for Advanced Study, where this work was begun, and INFN, the EU Project NANOQUESTFIT and the University of Trieste (FRA 2013) for financial support.  S.L.A acknowledges that final work on this paper was supported in part by National Science Foundation Grant No. PHYS-1066293 and the hospitality of the Aspen Center for Physics.



\begin{thebibliography}{99}
\bibitem{pik} I. Pikovski, M. Zych, F. Costa, and \v C Brukner, {\it Nature Physics} {\bf 11}, 668 (2015).
\bibitem{pik2} M. Zych, F. Costa, I. Pikvoski, and \v C Brukner, {\it Nature Commun.} {\bf 2}, 505 (2011).
\bibitem{bonder} Y. Bonder, E. Okon, and D. Sudarsky, ``Comment on `Universal decohrence due to gravitational time dilation'",
arXiv:1507.05320.
\bibitem{diosi} L. Di\'osi, ``Centre of mass decoherence due to time dilation: paradoxical frame-dependence'',
arXiv:1507.05828.
\bibitem{cow}  R. Colella, A. W. Overhauser, and S. A. Werner,  {\it Phys. Rev. Lett.} {\bf 34}, 1472 (1975); H. Rauch and
S. Werner, {\it Neutron Interferometry}, Oxford University Press (2000).
\bibitem{coll} E. Joos and H. D. Zeh, {\it Z. Phys. B: Condens. Matt.} {\bf 59}, 223 (1985);  L. Di\'osi, {\it Europhys. Lett.} {\bf 30}, 63 (1995); K. Hornberger and J. E. Sipe, {\it Phys. Rev.} A {\bf 68}, 012105 (2003); see
S. L. Adler,  {J. Phys. A: Math. Gen.} {\bf 39}, 14067 (2006) for a survey and further references.
\end{thebibliography}
\end{document}